\RequirePackage{lineno}
\documentclass[twocolumn,letterpaper,aps,prc,longbibliography,superscriptaddress,nofootinbib,floatfix]{revtex4-2}

\usepackage{graphicx}   
\usepackage{lipsum}
\usepackage{tabularx}
\usepackage{amsmath}
\usepackage[normalem]{ulem}
\usepackage{xspace}	
\usepackage{easyReview}
\usepackage{orcidlink}

\makeatletter
\newcommand{\fmarki}{*}
\newcommand{\fmarkii}{\ensuremath{\dagger}}
\def\@fnsymbol#1{{\ifcase#1\or \fmarki\or \fmarkii \else\@ctrerr\fi}}
\makeatother
\renewcommand{\fmarkii}{\&}

\newcommand{\pt}{\mbox{$p_{\perp}$}\xspace}

\newcommand{\ptmin}{\mbox{$p_{\perp\min}$}\xspace}
\newcommand{\ptmineff}{\mbox{$p_{\perp\min}^{\mathrm{eff}}$}\xspace}
\newcommand{\Dy}{\mbox{$\Delta y$}\xspace}

\newcommand{\shat}{\mbox{$\hat{s}$}\xspace}
\newcommand{\that}{\mbox{$\hat{t}$}\xspace}

\newcommand{\ATLAS} {ATLAS\xspace}

\newcommand{\LHAPDF} {{\textsc{LHAPDF}}\xspace}
\newcommand{\FASTJET} {{\textsc{fastjet}}\xspace}
\newcommand{\PYTHIAeight} {{\textsc{pythia8}}\xspace}

\newcommand{\NNPDFLO} {{\textsc{NNPDF31\_lo\_as0130}}\xspace}
\newcommand{\MSTWNLO} {{\textsc{MSTW2008nlo68cl}}\xspace}

\newcommand{\TeV}{\mbox{TeV}\xspace}
\newcommand{\GeV}{\mbox{GeV}\xspace}

\newcommand{\sqs}{\mbox{$\sqrt{s}$}\xspace}
\newcommand{\sqshat}{\mbox{$\sqrt{\hat{s}}$}\xspace}
\newcommand{\mjj}{\mbox{$M_{jj}$}\xspace}
\newcommand{\mgg}{\mbox{$M_{\gamma\gamma}$}\xspace}

\newcommand{\mll}{\mbox{$M_{ll}$}\xspace}
\newcommand{\mjjmin}{\mbox{$M_{jj\mathrm{min}}$}\xspace}
\newcommand{\md}{\mbox{$M_{D}$}\xspace}
\newcommand{\pp}{\mbox{$pp$}\xspace}

\begin{document}

\title{Dijets with a large rapidity separation in the next-to-leading order BFKL formalism for searches of large extra dimensions at colliders}

\newcommand{\pnpi}{Petersburg Nuclear Physics Institute, NRC Kurchatov Institute, Gatchina, 188300 Russia}

\affiliation{\pnpi}

\author{Anatolii~Iu.~Egorov\orcidlink{0000-0003-4936-6962}}\email{egorov\_aiu@pnpi.nrcki.ru} \affiliation{\pnpi} 
\author{Victor~T.~Kim\orcidlink{0000-0001-7161-2133}} \email{kim\_vt@pnpi.nrcki.ru} \affiliation{\pnpi} 
\author{Viktor~A.~Murzin\orcidlink{0000-0002-0554-4627}} \affiliation{\pnpi} 
\author{Vadim~A.~Oreshkin\orcidlink{0000-0003-4749-4995}} \affiliation{\pnpi} 
\date{\today}


\begin{abstract}

Search for the gravity with large extra dimensions at collider energies is considered in the trans-Planckian eikonal regime, i.e., when $\sqshat \gg \md \gg \sqrt{-\that}$. Here $\hat{s}$ and $\hat{t}$ are the Mandelstam variables of colliding parton-parton system and $M_D$ is the Planck mass scale in the space-time with compactified $n_D$ extra dimensions.  
A relevant observable for this regime may be the cross section of high-mass ($\mjj\sim\sqshat \gg \md$) dijet production with a large rapidity separation. 
Then the standard model  background should be calculated within the next-to-leading logarithmic (NLL) approximation of Lipatov-Fadin-Kuraev-Balitsky (BFKL) formalism of quantum chromodynamics (QCD) suitable for $\sqrt{\hat{s}}\gg\sqrt{-\hat{t}}\gg\Lambda_\mathrm{QCD}$. 
In this work the signal of the large extra dimension gravity as well as the NLL BFKL QCD background are estimated for the high-luminosity Large Hadron Collider (HL-LHC) and future colliders such as FCCpp and CEPC-SppC.


\end{abstract}

\maketitle

\section{INTRODUCTION}

Recent measurements of dijet production cross sections with a large rapidity separation, conducted by \ATLAS and CMS at the  Large Hadron Collider (LHC), revealed shortcomings in the commonly used hard regime of quantum chromodynamics (QCD), when describing the observed dijet cross sections at large rapidities~\cite{ATLAS:2011yyh, ATLAS:2014lzu, CMS:2012rfo, CMS:2012xfg, CMS:2016qng, CMS:2017oyi, TOTEM:2021rix, CMS:2021maw, Egorov:2022msn, Egorov:2023duz, Egorov:2025avv}. In the hard regime ($\sqrt{s} \simeq Q \gg \Lambda_{\mathrm{QCD}}$), where the characteristic hard scale of the collision, $Q$, is of the order of the collision energy, $\sqrt{s}$,  the well-established Gribov-Lipatov-Altarelli-Parisi-Dokshitzer (DGLAP) equations are successful to describe logarithmic 
$Q$-evolution \cite{Gribov:1972ri, Gribov:1972rt, Lipatov:1974qm, Altarelli:1977zs, Dokshitzer:1977sg}. In fact, with increasing of collision energy the measurement kinematics tends to the domain of the semi-hard QCD regime  ($\sqrt{s} \gg Q \gg \Lambda_{\mathrm{QCD}}$), where the logarithmic $s$-evolution is governed by the Lipatov-Fadin-Kuraev-Balitsky (BFKL) equation \cite{Kuraev:1976ge, Kuraev:1977fs, Balitsky:1978ic}. This implies that when potential signals of new physics are expected to dominate in large rapidity regions with semi-hard kinematics, the standard model (SM) QCD background should be calculated incorporating BFKL evolution. Note, that
BFKL evolution in the next-to-leading logarithmic (NLL) approximation~\cite{Fadin:1998py, Ciafaloni:1998gs, Egorov:2023duz} includes also the leading DGLAP $Q$-terms.  

Recently, it was shown that cross section of dijet production with a large rapidity separation at the LHC
\cite{CMS:2021maw} can be described by the NLL BFKL evolution
\cite{Egorov:2023duz}, while DGLAP prediction significantly overestimates the data. 
Therefore, searches for new physics beyond the SM at the high-luminosity Large Hadron Collider (HL-LHC)~\cite{Aberle:2749422} and future colliders, such as FCCpp~\cite{FCC:2018vvp}  and CEPC-SppC~\cite{Gao:2021bam}, by dijets with a large rapidity separation should take into account new BFKL evolution dynamics for SM background estimates.

One example of new physics beyond the SM that can enhance dijet production with a large rapidity separation is gravity with large compactified extra dimensions \cite{Arkani-Hamed:1998jmv} in the trans-Planckian eikonal regime  \cite{Giudice:1998ck, Giudice:2001ce, Kim:2010zzf, Kim:2011zzj}. The model of gravity with large compactified extra dimensions was proposed by Arkani-Hamed, Dimopoulos, and Dvali (ADD) \cite{Arkani-Hamed:1998jmv}  to address the hierarchy puzzle between the observable scales of the SM and gravity.

The ADD gravity model was investigated at the LHC in proton-proton (\pp) collisions with $\sqs=8$ and $13$~\TeV in dilepton, diphoton, and dijet final states, focusing on the regime where the parton collision energy, $\sqshat\sim\sqrt{-\that}$, is below the assumed Planck scale in the presence of large extra dimensions, $\md$ \cite{ATLAS:2014gys, CMS:2014lcz, CMS:2018ipm, CMS:2018dqv, CMS:2018ucw, CMS:2018nlk, CMS:2021ctt}. Here, $\hat{s}$ and $\hat{t}$ denote the Mandelstam variables for the parton-parton collisions.  A lower bound on the gravity scale in the ADD model was established by the measurements at approximately $\md \sim 7\div10$~\TeV, depending on the specific model convention. These searches are based on the effective linearized gravity framework, formulated by  Giudice, Rattazzi and Wells (GRW)~\cite{Giudice:1998ck}, which leads to divergent tree-level diagrams due to integrating over Kaluza-Klein (KK) modes in the compactified extra dimensions. The ultraviolet cutoff is taken to be $\md$.  Naive dimensional analysis and unitarity considerations suggest that the effective ADD gravity theory  may remain valid up to energies near or somewhat above the \md scale \cite{Giudice:1998ck}. In general, effective theories are expected to apply at energies much lower than the new physics scale, i.e., $\sqshat \ll \md$. Therefore, when $\sqshat \sim \sqrt{-\that} \sim \md$, as is the case in the LHC searches, interpreting the results requires extra caution.

The trans-Planckian eikonal regime, introduced by GRW in Ref.~\cite{Giudice:2001ce}, is an alternative kinematic scenario for probing ADD gravity. In this regime $\sqshat \gg \md$, while the momentum transferred by the exchanged $t$-channel gravitons satisfies $\sqrt{-\that} \ll \md$. The eikonalization process removes ultraviolet sensitivity \cite{Giudice:2001ce}, because it favors gravitons with large wavelengths, which are insensitive to ultraviolet local counterterms. Partons scattered in this regime produce dijets that are widely separated in rapidity. Consequently, the dominant SM background for ADD gravity searches in the trans-Planckian eikonal kinematics comes from QCD interactions in the semi-hard regime.

In this work, we present the results of the dijet production cross section calculation due to large extra dimension gravity in the ADD model in the trans-Planckian eikonal regime and compare them with QCD background estimates performed using two approaches, accounting separately for DGLAP or BFKL evolution. The calculations are conducted under conditions expected at the HL-LHC and future colliders, such as FCCpp and CEPC-SppC, with collision energies up to 100 TeV. The calculations indicate that applying DGLAP dynamics at large rapidities in the semi-hard regime can cause an overestimation of the QCD background by several orders of magnitude, potentially leading to misinterpretation of experimental data and a missed chance to identify new physics.

\section{Gravity in space-time with large compactified extra dimensions}
\label{sec:ADDgrav}

The concept behind the ADD model which explains the observed weakness of gravity relative to other fundamental forces is the following~\cite{Arkani-Hamed:1998jmv}: the graviton is expected to propagate through the usual four-dimensional space-time as well as $n_D$ extra spatial dimensions, whereas the other fundamental fields are confined to the four-dimensional space-time submanifold. The gravitational potential between two masses in the presence of $n_D$ extra dimensions behaves as

\begin{equation}\label{eq:vpot}
V(r)\sim\frac{1}{M_D^{2+n_D}} \frac{1}{r^{1+n_D}},
\end{equation}

where $M_D$ is the fundamental parameter that sets the characteristic scale of gravity in the $4+n_D$-dimensional space-time. If the $n_D$ extra dimensions are compactified with size $R$, this modified gravitational potential behavior can be observed only at short distances where $r\ll R$. At large distances $r\gg R$, the gravitational potential is expected to follow the ordinary $1/r$ behavior

\begin{equation}\label{eq:vpotlarge}
V(r)\sim\frac{1}{M_D^{2+n_D}R^{n_D}} \frac{1}{r}.
\end{equation}

This immediately shows that the large magnitude of the Planck mass $M_{\mathrm{Pl}}$ is determined by the volume of the extra dimensions, such that $M_{\mathrm{Pl}}=M_D^{1 + n_D/2}R^{n_D/2}$. Assuming that the gravity scale $M_D$ is on the order of the electroweak scale, $m_{\mathrm{EW}}$, the required size $R$ of the extra dimensions can be determined as a function of their number $n_D$

\begin{equation}\label{eq:dsize}
R\sim 10 ^{\frac{30}{n_D}-17}\mathrm{cm}\times\bigg(\frac{1\mathrm{TeV}}{m_\mathrm{EW}}\bigg)^{1 + \frac{2}{n_D}}.
\end{equation}

While gravity with $n_D=1$ extra dimension is excluded due to the fact that the $1/r$ gravitational behavior has been tested down to submillimeter distances and shows no deviation, scenarios with $n_D>1$ extra dimensions remain viable and are still being actively explored. 

If $M_D\sim 1$~TeV, then the trans-Planckian regime, where $\sqrt{s} \gg M_D$, can be accessed at the HL-LHC as well as at future colliders such as FCCpp and CEPC-SppC. The propagation of gravitons through the compact extra dimensions is described by KK excitations. Due to the large size $R$ of the extra dimensions, the masses of the KK modes form an almost continuous spectrum. The virtual exchange of these states would appear as a broad excess in cross sections above the SM background.

The work by GRW Ref.~\cite{Giudice:2001ce} explores the possibility of parton scattering through multiple KK graviton exchanges in the trans-Planckian regime using the eikonal approximation, where $\sqshat\gg\md \gg \sqrt{-\that}$. In this kinematic regime, the gravity signal is expected to manifest itself as an excess in the production of dijets with jets which are widely separated in rapidity. This approach resums an infinite series of graviton exchanges, valid when the momentum transfer is small compared to the center-of-mass energy. The summation of an infinite series of successive graviton exchanges results in the following expression for the parton-parton cross section:
\begin{equation}\label{eq:add_grw_xs}
\frac{d\hat{\sigma}_{\mathrm{eik}}}{d\Delta y} = \frac{\pi b_c^4\hat{s}e^{\Delta y}}{(1 + e^{\Delta y})^2}\big|F_D(x)|^2,
\end{equation}
where $b_c$ is a length scale defining the upper boundary of classical gravity consideration. The function $F_D(x)$ is the essential component of the eikonal amplitude. The definitions of $b_c$ and $F_D(x)$ are given by Eqs. (18) and (20) of Ref. \cite{Giudice:2001ce}; $x=b_c\sqrt{\hat{s}}/\sqrt{1+e^{\Delta y}}$; and $\Delta y$ is the separation in rapidity between the two jets of a dijet. In the trans-Planckian ($\sqrt{\hat{s}}/M_D\gg 1$) eikonal ($-\hat{t}/\hat{s}\ll 1$) regime $\sqrt{\hat{s}}$ is equal to jet-jet invariant mass $M_{jj} = \sqrt{\hat{s}}$, and $-\hat{t}/\hat{s}=1/(1+e^{\Delta y})$. Using the parton-parton cross section equation~(\ref{eq:add_grw_xs}), one can convolute it with parton distribution functions (PDFs) to obtain the dijet production cross section in \pp collisions due to multiple graviton exchanges. It is important to note that the cross section equation~(\ref{eq:add_grw_xs}) is valid only in the regime where $-\hat{t}/\hat{s}\ll1$, corresponding to large rapidity intervals ($\Delta y\gg 1$). How this expression matches the region of small $\Delta y$ remains basically unclear.

From the signature of the expected signal (high-mass dijets with a large rapidity separation) it is clear that the dominant SM background arises from QCD dijet production. Accurately modeling and estimating QCD background is essential for identifying any excess due to new physics.

\section{QCD background estimation}
\label{sec:QCDbackraund}

Recent results from the CMS experiment at the LHC report the first measurement of the cross section for Mueller-Navelet (MN) dijet production at TeV-scale energies \cite{CMS:2021maw}. MN dijets refer to the pair of jets in an event that have the largest separation in rapidity among all jets with transverse momentum above a given threshold $\ptmin$~\cite{Mueller:1986ey}. The CMS Collaboration also provided comparisons of their measurements with theoretical calculations that incorporate leading order (LO) and next-to-leading order (NLO) QCD matrix elements, enhanced by leading logarithmic (LL) DGLAP evolution of the parton shower. The comparison revealed that theoretical models using DGLAP evolution significantly overestimate the measured dijet production cross sections, by up to a factor of $6$ in the region of large rapidity separation $\Dy$. Noticeable deviations of DGLAP-based calculations from the measurements begin to appear for rapidity separations $\Dy \gtrsim 4$, indicating that the theoretical models based on the DGLAP evolution, become less accurate in this large rapidity interval region. 

The accuracy of LL BFKL-based calculations is generally insufficient, as the LL approximation tends to predict stronger effects than those observed in experimental measurements. This overestimation arises due to limitations inherent in the LL BFKL approach, such as neglecting certain higher-order corrections like running of the strong coupling and kinematical constraints such as energy conservation. Consequently, NLL corrections~\cite{Fadin:1998py, Ciafaloni:1998gs} are necessary to achieve a more accurate and realistic description of the data. The NLL BFKL calculations rely on three key components: the process-independent BFKL Green's function, process-dependent impact factors, and a procedure to resolve the QCD renormalization scale ambiguity. The detailed calculations for the BFKL Green's function and impact factors for MN dijets can be found in Refs~\cite{Caporale:2015uva, Ducloue:2013hia}. The approach that systematically eliminates scale-setting ambiguities was developed by Brodsky, Fadin, Kim, Lipatov and Pivovarov (BFKLP)~\cite{Brodsky:1998kn}. This method essentially generalizes the Brodsky, Lepage, and Mackenzie (BLM) optimal scale-setting procedure~\cite{Brodsky:1982gc} to the non-Abelian case.

Calculations performed in Ref. \cite{Egorov:2023duz}, based on the NLL BFKL evolution, demonstrate good agreement with the measurement data in the region $\Dy>4$. This provides new evidence supporting the possible manifestation of BFKL evolution effects at TeV energy scales and suggests that the QCD background is more accurately described by the NLL BFKL evolution in the regime of large $\Dy$. 

In this work, we provide an estimation of the QCD background based on both DGLAP and BFKL evolution. The DGLAP-based estimation is performed as a convolution of PDFs evolving according to the LL DGLAP equations with LO QCD matrix elements. Additionally, a DGLAP-based prediction is generated using the Monte Carlo (MC) event generator package \PYTHIAeight~\cite{Sjostrand:2007gs}.  The BFKL-based predictions are supplied at both LL and NLL accuracy. The BFKL calculations follow the methodology described in Refs.~\cite{Caporale:2015uva, Caporale:2012ih, Egorov:2023duz, Celiberto:2016ygs}. The BFKLP scales were precalculated exactly using the equation (40) from Ref.~\cite{Caporale:2015uva} as a function of $\sqshat$ and transverse momenta of jets forming a dijet system $p_{\perp 1}$ and $p_{\perp 2}$.

\section{Definition of observable}
\label{sec:Def}

To search for ADD gravity in the trans-Planckian eikonal regime using dijet final states, one must select MN dijets and apply the high-mass cut on the dijet mass, $\mjj > \mjjmin$. The selection procedure is as follows: first, select all jets in the event with transverse momentum above \ptmin. Next, identify the pair of jets with the largest rapidity separation, known as the MN dijet. Finally, from these MN dijets, choose those with $\mjj > \mjjmin$. In this work we set $\ptmin = 20$~\GeV. We select only MN dijets formed of jets with rapidity $|y|<y_{\max} = 4.7$, to match the geometry of the detectors at the LHC. This selection is largely based on the criteria from previous CMS studies~\cite{CMS:2012xfg,CMS:2016qng,CMS:2021maw}, with the difference that \ptmin is set to $20$~GeV instead of $35$~\GeV. This lower \ptmin value is experimentally reasonable and allows for selecting more dijet events with a large rapidity separation, \Dy. However, it should be noted that applying the high dijet mass cut at given \Dy effectively selects dijets composed of jets with higher 
\pt. 

In the definition of the observable given above, we opted for MN jets instead of inclusive jets due to several reasons. 

First, the MN dijet cross section is much easier to compute within the NLL BFKL framework than, for example, the inclusive dijet cross section. Here the inclusive dijet production includes all pairwise jet combinations meeting selection criteria. As shown in Refs. \cite{CMS:2016qng} and \cite{CMS:2021maw}, NLL BFKL calculation better reproduces data at large \Dy than other QCD approaches. 

Second, at large \Dy, the MN dijet cross section approaches the inclusive one. At large \Dy and \ptmin thresholds, little energy remains for a third high-\pt jet with large rapidity. This is even more true when large \mjj selections are imposed. In case of large \mjj selection nearly all collision energy is spend for the MN dijet production. Additionally, within the DGLAP evolution, leading dijet (leading dijet consists of two jets with highest \pt in the event) cross sections are more natural to calculate. At large \Dy and high \mjj, the MN dijet is almost surely the leading one. Thus, differences between MN, inclusive, and leading dijet cross sections should vanish in this regime.

Third, it is known, the BFKL dynamic produces azimuthal angular decorrelations of jets forming MN dijet due to real QCD emission of additional jets within the rapidity interval between jets~\cite{DelDuca:1993mn, SabioVera:2007ndx}. We expect transverse momentum angular correlations of current calculation [employing Eq.~(\ref{eq:add_grw_xs})] for high-\mjj dijets at large \Dy produced by $t$-channel multiple graviton exchange within ADD gravity  to resemble the evolution of DGLAP more than the BFKL one. This is because, the partonic cross section in Eq. (\ref{eq:add_grw_xs}) omits real graviton emission, so there is no azimuthal decorrelation at the subprocess level. Observing ADD-induced azimuthal decorrelations would require including real graviton emission alongside the sequential graviton exchanges accounted in Eq. (\ref{eq:add_grw_xs}). This approach, to our knowledge, is not yet successfully implemented.  If real graviton emission is included in the calculation in the trans-Planckian eikonal regime, it can produce non-negligible azimuthal decorrelation. From this point of view, the azimuthal decorrelation of jets forming high-mass, large-\Dy dijets is interesting to measure.

Finally, the measured MN dijet cross section shows more pronounced deviations from the DGLAP predictions [Ref.~\cite{CMS:2021maw}] than measured azimuthal decorrelations [Ref.~\cite{CMS:2016qng}]. Specifically, the DGLAP predictions overestimates the measured MN dijet cross section by a factor of 5 at $\Dy\sim7.5$ for $\sqrt{s}=2.76$ TeV [Ref.~\cite{CMS:2021maw}], while the deviation of the DGLAP predictions from measured azimuthal decorrelation at  $\sqrt{s}=7$ TeV is comparable to experimental systematic uncertainty [Ref.~\cite{CMS:2016qng}].

This study examines \pp collisions at energies of $\sqs=13$, $40$, and $100$~\TeV, spanning from current LHC energies up to those expected for FCCpp and CEPC-SppC. Although the ADD signal to QCD background ratio is expected to improve with increasing \mjjmin, the cross sections diminish significantly at large \mjjmin. Thus, the high dijet mass cuts are selected based on the projected integrated luminosities of the HL-LHC, FCCpp and CEPC-SppC. Specifically, for $\sqs=13$~\TeV we consider $\mjjmin=6$, and $9$~\TeV; for $\sqs=40$~\TeV, $\mjjmin=9$, and $30$~\TeV; and for $\sqs=100$~\TeV, $\mjjmin=30$, and $70$~\TeV.

\begin{figure*}[!ht]
\includegraphics[width=0.99\linewidth]{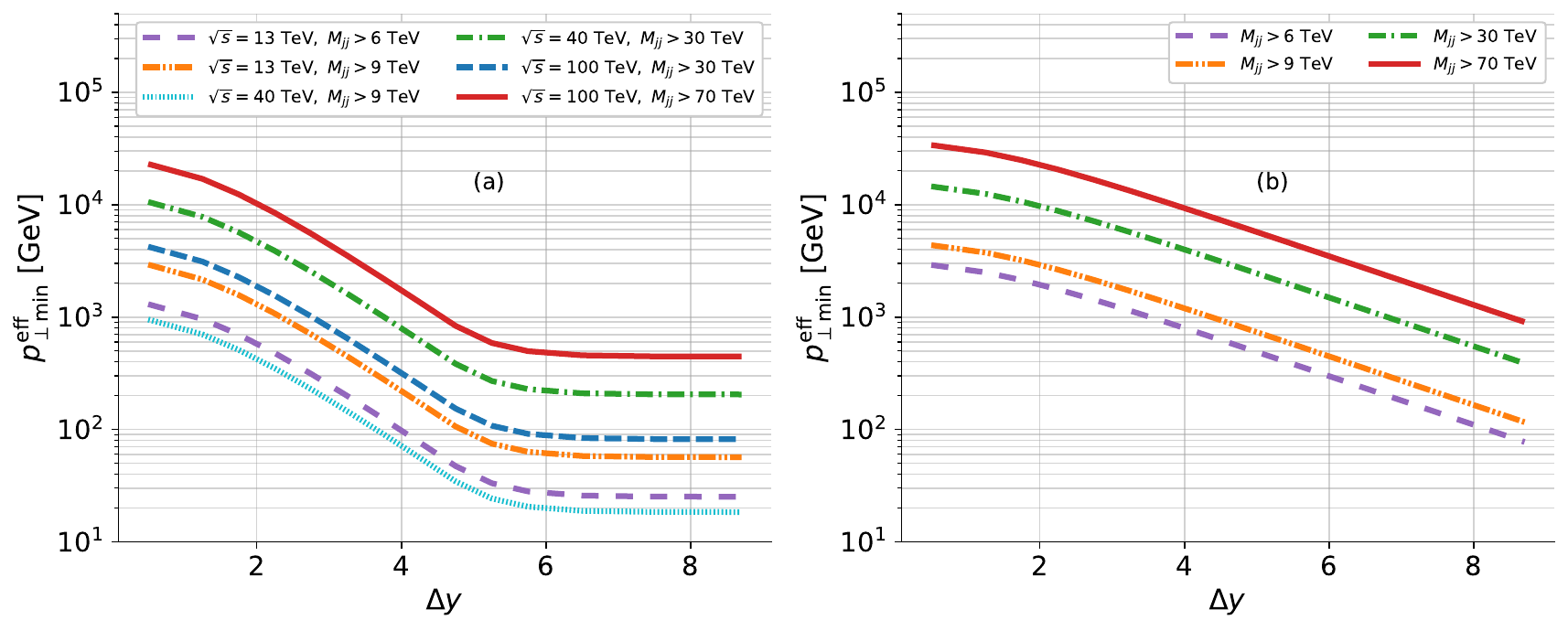}
\caption{Effective minimum transverse momentum, \ptmineff, of jets forming an MN dijet after applying the 
\mjj selection for the case $p_{\perp1}\neq p_{\perp2}$ (a); and for the case $p_{\perp1} = p_{\perp2}$ (b).
}
\label{fig:pt_min}
\end{figure*}

The dijet mass can be determined with the following expression:
\begin{equation}\label{eq:mjj}
M_{jj}^2 = 2p_{\perp1}p_{\perp2}(\cosh{(\Dy)}-\cos(\Delta \phi)),
\end{equation}
where $\Delta \phi$ represents the difference in azimuthal angle between the two jets forming the dijet. From here we see that imposition high \mjjmin selection on a dijet effectively select jets with \pt higher than \ptmin. In the BFKL kinematic regime, $p_{\perp1}$ can differ from $p_{\perp2}$ due to transverse momentum diffusion. The effective minimum jet transverse momentum, \ptmineff, is determined by \mjjmin, \Dy and $p_{\perp\max} = \sqs/(2\cosh(\max[-y_{\max}+\Dy, 0]))$. For the selections analyzed in this study, the effective minimum jet \pt in BFKL kinematics is shown in  Fig~\ref{fig:pt_min}(a).  It is observed that $\ptmineff<\ptmin=20$~\GeV only for the $\sqs=40$~\TeV with $\mjj>9$~TeV selection, where $\ptmineff\approx 18$~\GeV for $\Dy > 6$.  However, cases with a large imbalance, i.e. when the \pt of one jet is much smaller than that of the other, lead to an oscillating integrand in the BFKL calculation, causing these configurations to be suppressed relative to those with $p_{\perp1}\approx p_{\perp2}$. When $p_{\perp1} = p_{\perp2}$, as in the DGLAP kinematics, the effective minimum transverse momentum, \ptmineff, is determined by \mjjmin and \Dy. The corresponding \ptmineff values are presented in Fig~\ref{fig:pt_min}(b). It is evident that \ptmineff is significantly higher than $\ptmin=20$~\GeV in this case.

\label{sec:Num}
\section{Numerical calculations and systematic uncertainty}

The ADD gravity signal cross section is computed by convolving the partonic cross section given in Eq.~(\ref{eq:add_grw_xs}) with PDFs from \NNPDFLO~\cite{NNPDF:2017mvq}, accessed through the \LHAPDF framework~\cite{Buckley:2014ana}. The ADD model depends on two parameters: \md and the number of extra dimensions $n_D$. This study considers two values, $n_D = 2$ and $6$, covering the typical range. The values of Planck scale with extra dimensions, \md, are chosen so that the signal cross section is roughly comparable to the QCD background at $\Dy = 8.7$. The cross sections vanish near the kinematic limit $\Dy = 9.4$, imposed by the rapidity cut $|y|<y_{\max}=4.7$. Therefore, $\Dy = 8.7$ is chosen as the largest \Dy unaffected by the $|y|<y_{\max}=4.7$ selection. Since the partonic cross section in Eq.~(\ref{eq:add_grw_xs}) relies on the eikonal approximation with $\shat\gg|\that|$, signal calculations are performed for $\Dy > 3.25$ ($|\that|/\shat < 0.04$). At smaller \Dy values, the eikonal approximation may become less reliable. Therefore, the most accurate results using this approximation occur at the largest $\Dy = 8.7$ ($|\that|/\shat \approx 1.7 \cdot 10^{-4}$). Another way to estimate the lower bound for the trans-Planckian eikonal regime's applicability is to consider the onset of black hole (BH) formation. A rough criterion is when the impact parameter becomes equal the Schwarzschild radius, $R_{\mathrm{S}}$, for the system of mass \mjj, i.e.,  $1/\sqrt{-\hat{t}}=R_\mathrm{S}$. This yields the following lower bound on \Dy:

\begin{equation}\label{eq:Dymin}
\Delta y > \log\big((R_\mathrm{S}M_{jj})^2 - 1\big).
\end{equation}

In the ADD gravity signal calculation, \mjj is integrated from \mjjmin  to \sqs. Strictly speaking, at fixed \Dy, BH first forms when $\mjj = \sqs$. However, this is suppressed by the sharp drop in structure functions as $x\rightarrow1$. For a rough estimate of the lower \Dy bound, we take $\mjj = \sqs$. Higher $M_D$ at fixed \mjj leads to lower \Dy bound, because gravity becomes weaker, and visa versa. As we are about to see, the most interesting results are for $M_D = 3$~TeV at $\sqs = 13$~TeV; $M_D = 10$~TeV at $\sqs = 40$~TeV; and  $M_D = 20$~TeV at $\sqs = 100$~TeV. For these values $\Dy_{\min}$ is found between 3.5 and 5.5 according to Eq. (\ref{eq:Dymin}). For \Dy below $\Dy_{\min}$, some signal leakage due to BH formation may occur. Nevertheless, we believe a step toward larger \Dy values is needed for Eq. (\ref{eq:add_grw_xs}) to be fully reliable. Thus, we base our conclusions on ADD gravity sensitivity in the trans-Planckian eikonal regime on background-signal comparisons at $\Dy\sim8.7$. If future measurements reveal an excess over the SM QCD background at lower \Dy, it could still indicate ADD gravity with $M_D$ values larger than those considered in our study.

The QCD background within the DGLAP approximation is evaluated by either convolving LO QCD partonic cross sections with \NNPDFLO PDFs evolved through LL DGLAP evolution or using the \PYTHIAeight MC generator with the CP5 tune~\cite{CMS:2019csb}. Originally \PYTHIAeight performed at LO+LL DGLAP accuracy as well. However, the CP5 tune updates it by using a NLO strong coupling constant and NNLO PDFs, along with rapidity ordering in the initial-state radiation. This rapidity ordering makes behavior of \PYTHIAeight predictions closer to the BFKL kinematics, improving the modeling of gluon emissions and parton evolution for high-energy collisions. The validity of these adjustments require careful theoretical consistency checks. We developed a custom \PYTHIAeight plugin to improve computational efficiency in MC simulations by biasing event generation toward high dijet masses at the parton level and correcting this bias with compensating weights to maintain accurate overall distributions. The plugin uses dijet mass at parton level, and its performance was validated against unweighted simulations. The applied exponential weight-bias function enhances computational efficiency of high dijet mass event generation while keeping weights bounded and uncertainties manageable. The form of the bias function is
\begin{equation}\label{eq:hook}
\exp\left(\frac{\sqrt{\hat{s}}}{a}\right) - 1 + \frac{b}{a},
\end{equation}
where the parameters $a$ and $b$ are chosen to optimize the generation efficiency. Jets in the \PYTHIAeight calculations are reconstructed using the anti-$k_T$ algorithm~\cite{Cacciari:2008gp} implemented in the \FASTJET package~\cite{Cacciari:2011ma}, with a jet size parameter of 0.4, consistent with standard practice in LHC measurements at $\sqs=13$~\TeV.

The QCD background under the BFKL approximation is computed as outlined in Sec.~\ref{sec:QCDbackraund}, providing both LL and NLL BFKL predictions. For the LL BFKL approximation, \NNPDFLO PDFs are employed, while the NLL BFKL calculations use \MSTWNLO~\cite{Martin:2009iq} PDFs. In the NLL BFKL corrections, the $k_T$ algorithm of jet reconstruction is applied with a jet size of 0.4, following the procedure detailed in Ref.~\cite{Colferai:2015zfa}. 

The systematic uncertainty of the NLL BFKL calculations are estimated. As previously noted, NLL BFKL calculations depend strongly on renormalization and factorization scales. The BFKLP procedure following the idea of restoring conformal properties of high-energy asymptotic regime fixes optimal value of renormalization and factorization scales tightly. To estimate the scale uncertainty within BFKL calculations it is possible to vary $s_0$ parameter, which define the beginning of the high-energy asymptotic. The central value of $s_0$ is typically defined by the product $|p_{\perp1}|\times |p_{\perp 2}|$. It is
varied by factors 2 and 0.5 to obtain the uncertainty. 

A further source of systematic uncertainty in these calculations stems from incomplete knowledge of the PDFs. To estimate this uncertainty, we use MC replicas from the PDF4LHC15\_NLO\_MC set \cite{Butterworth:2015oua}. This particular PDF set provides a combination of three PDF sets, namely CT14 \cite{Dulat:2015mca}, MMHT2014 \cite{Harland-Lang:2014zoa}, and NNPDF30 \cite{NNPDF:2014otw}, and properly accounts for the uncertainties from each of these sets at NLO accuracy.

The total systematic uncertainty of the NLL BFKL calculations combines these contributions in quadrature. When \mjjmin is not too close to \sqs, both sources contribute comparably. However, as \mjjmin approaches \sqs, the PDF uncertainty dominates due to its growth in the $x\rightarrow 1$ limit.

\section{Results and discussion}
\label{sec:Results}

\begin{figure*}[!ht]
\includegraphics[width=0.99\linewidth]{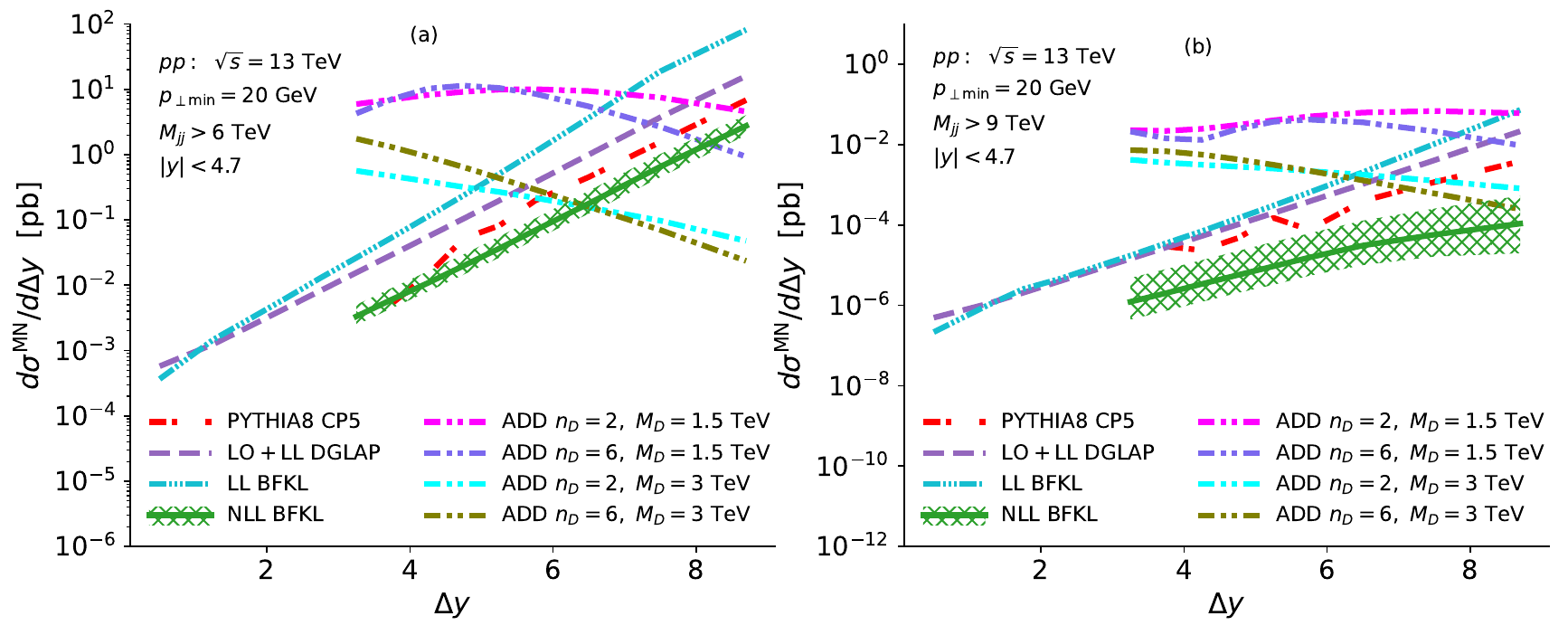}
\caption{The MN dijet cross section with the imposed dijet mass selection $\mjj>\mjjmin$ in \pp collisions at $\sqs=13$~\TeV. The ADD gravity signal is shown for various parameter choices, including the number of extra dimensions $n_D$, and the Plank scale $\md$ in the presence of $n_D$ extra dimensions. The QCD background includes either contributions calculated with the LO+LL DGLAP or LL/NLL BFKL corrections. Panel (a) corresponds to $\mjjmin=6$~\TeV, and panel (b) to $\mjjmin=9$~\TeV.
}
\label{fig:13TeV}
\end{figure*}

\begin{figure*}[!ht]
\includegraphics[width=0.99\linewidth]{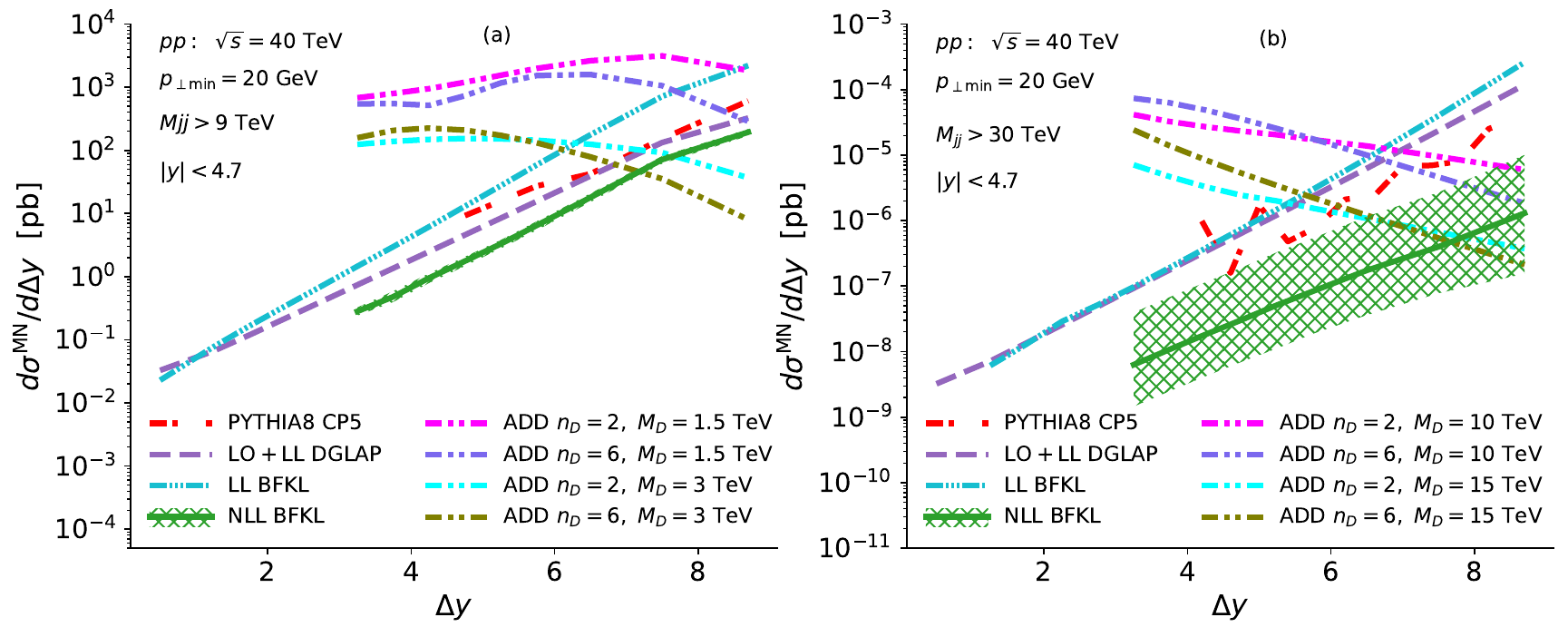}
\caption{The MN dijet cross section with the imposed dijet mass selection $\mjj>\mjjmin$ in \pp collisions at $\sqs=40$~\TeV. The ADD gravity signal is shown for various parameter choices, including the number of extra dimensions $n_D$, and the Plank scale $\md$ in the presence of $n_D$ extra dimensions. The QCD background includes either contributions calculated with the LO+LL DGLAP or LL/NLL BFKL corrections. Panel (a) corresponds to $\mjjmin=9$~\TeV, and panel (b) to $\mjjmin=30$~\TeV.
}
\label{fig:40TeV}
\end{figure*}

\begin{figure*}[!ht]
\includegraphics[width=0.99\linewidth]{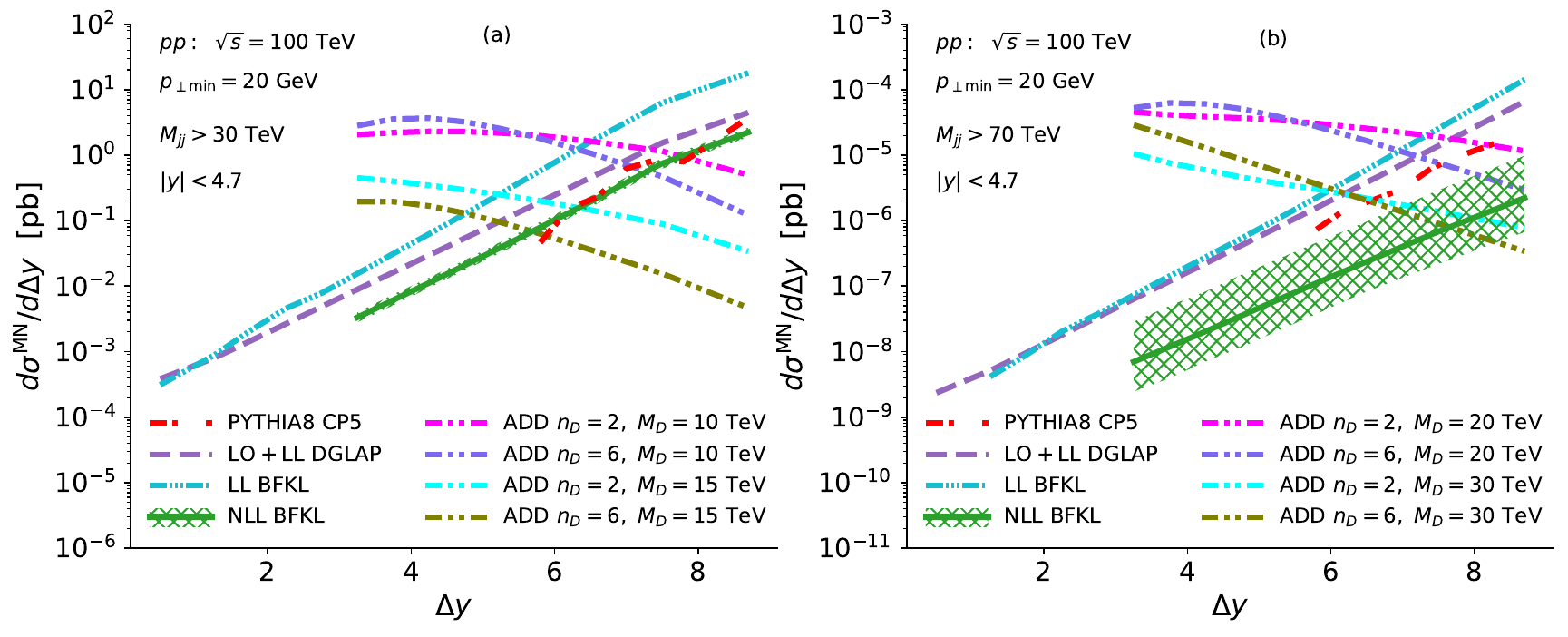}
\caption{The MN dijet cross section with the imposed dijet mass selection $\mjj>\mjjmin$ in \pp collisions at $\sqs=100$~\TeV. The ADD gravity signal is shown for various parameter choices, including the number of extra dimensions $n_D$, and the Plank scale $\md$ in the presence of $n_D$ extra dimensions. The QCD background includes either contributions calculated with the LO+LL DGLAP or LL/NLL BFKL corrections. The systematic uncertainty of the NLL BFKL calculation is denoted by the hatched green band. Panel (a) corresponds to $\mjjmin=30$~\TeV, and panel (b) to $\mjjmin=70$~\TeV.
}
\label{fig:100TeV}
\end{figure*}

The results of the MN dijet cross section calculations with the high dijet mass selection $\mjj>\mjjmin$ in \pp collisions at $\sqs=13$, $40$ and $100$~\TeV are shown in Figs.~\ref{fig:13TeV}, \ref{fig:40TeV}, and \ref{fig:100TeV}, respectively.

The signal calculation for gravity with large extra dimensions, as proposed by the ADD model, is presented for several parameter choices. The number of extra dimensions considered are $n_D = 2$ and $6$, reflecting typical values studied in this theoretical framework. This allows exploration of how the cross sections and phenomenology depend on the dimensionality of the extra spatial dimensions. The examined values of the Planck scale in the presence of extra dimensions are as follows: $\md = 1.5$ and $3$~\TeV for $\sqs = 13$~\TeV with $\mjj> 6$ and $9$~\TeV [Fig.~\ref{fig:13TeV}]; $\md = 1.5$ and $3$~\TeV for $\sqs = 40$~\TeV with $\mjj> 9$~\TeV [Fig.~\ref{fig:40TeV}(a)]; $\md = 10$ and $15$~\TeV for $\sqs = 40$~\TeV with $\mjj> 30$~\TeV [Fig.~\ref{fig:40TeV}(b)];  $\md = 10$ and $15$~\TeV for $\sqs = 100$~\TeV with $\mjj> 30$~\TeV [Fig.~\ref{fig:100TeV}(a)]; $\md = 20$ and $30$~\TeV for $\sqs = 100$~\TeV with $\mjj> 70$~\TeV [Fig.~\ref{fig:100TeV}(b)]. 

The predictions of the ADD model exhibit less sensitivity to the number of extra dimensions $n_D$ and are more strongly impacted by the value of the scale 
\md. The reduced sensitivity of ADD model predictions to the number of extra dimensions $n_D$ can be understood as follows: the change of $n_D$ at a fixed scale \md alters the size of the extra dimensions, which remain large enough for the particles to be insensitive to this change thanks to the vast difference between particle interaction scales and the scale corresponding to the size of extra dimensions. In contrast, varying \md directly modifies the strength of the gravitational coupling to matter, thus having a more pronounced effect on observable signals.

The QCD background predictions in various approximations are provided with: LO+LL DGLAP approximation; DGLAP-based MC generator \PYTHIAeight with the CP5 tune; and calculations including LL and NLL BFKL corrections. All of them are presented for all considered collision energies and dijet mass selections. This provides a comprehensive comparison of different theoretical approaches and modeling techniques for the QCD background across a variety of kinematic conditions. The NLL BFKL predictions consistently yield lower estimates for the QCD background compared to other calculations (up to 2 orders of magnitude). Additionally, the high dijet mass selection $\mjj>\mjjmin$ has the most pronounced effect on the NLL BFKL predictions among the various QCD approaches, leading to a stronger suppression of the background in this regime. 

The systematic uncertainty of NLL BFKL calculation is shown as a hatched green band in Figs.~\ref{fig:13TeV}, \ref{fig:40TeV}, and \ref{fig:100TeV}. For the highest \mjjmin selections (the most sensitive regions for ADD gravity searches in the trans-Planckian eikonal regime) the total uncertainty is about a factor of ~5 (or ~1/5) across all collision energies. This remains smaller than the difference between DGLAP and NLL BFKL background estimates.

Calculations incorporating the full NLL BFKL resummation provide a better description of dijet measurements at large rapidity separations \Dy in the LHC experiments at \TeV scales.~\cite{ATLAS:2011yyh, ATLAS:2014lzu, CMS:2012rfo, CMS:2012xfg, CMS:2016qng, CMS:2017oyi, TOTEM:2021rix, CMS:2021maw, Egorov:2022msn, Egorov:2023duz, Egorov:2025avv}. Given that, the NLL BFKL predictions are regarded as the most reliable estimations for the QCD background in this kinematic regime.

Considering the largest rapidity separation currently accessible in experiments at the LHC $\Dy=8.7$ and the background calculated including NLL BFKL corrections, it can be concluded that measurements of MN dijet production with the high dijet mass selection in \pp collisions are sensitive to ADD gravity for $\md<3$~\TeV at $\sqs = 13$~\TeV; for $\md<10$~\TeV at $\sqs = 40$~\TeV; and for $\md<20$~\TeV at $\sqs = 100$~\TeV. In case the ADD signal is comparable to the QCD background, the large PDF systematic uncertainty can prevent effective separation of the two. More precise PDF measurements at large $x$ are therefore highly desirable. The sensitivity of measurements in trans-Planckin eikonal regime to the ADD gravity may improve further if the transition to eikonal regime occurs at lower values of \Dy.

The estimated sensitivity of the proposed approach appears to be lower compared to existing searches conducted at the LHC~\cite{ATLAS:2014gys, CMS:2014lcz, CMS:2018ipm, CMS:2018dqv, CMS:2018ucw, CMS:2018nlk, CMS:2021ctt}, which establish a lower bound on the scale of ADD gravity \md at $7\div10$~\TeV. However, it is important to remember that the LHC searches rely on divergent behavior of effective theory near the scale of new physics. The masses of observed dilepton (diphoton) systems reach 3(2.6)~TeV which are close to the established lower bound of \md.  The complementary approach discussed here explores a distinct kinematic regime. In trans-Planckian eikonal regime the scale of gravitons is defined by the $\that$ variable, which at $\Delta y = 8.7$ is $\sim1\%$ of \mjj. Thus, the scale of gravitons is about 90 GeV, when $\mjj=9$ TeV. This scale of gravitons is much lower then considered \md = 3 TeV at 13~\TeV collision energy. Studying gravity in the trans-Planckian eikonal regime therefore will help to reinforce and validate the conclusions of existing searches.

An increase in high-mass dijet production at large \Dy inevitably leads to the simultaneous enhancement of the high-\pt forward jet spectrum (here, forward denotes large rapidity region). However, calculating single-jet spectra within the BFKL approach remains challenging. Therefore, further development of BFKL calculation methods is needed in search of an ADD gravity signal in forward jet \pt spectra.

High-mass dilepton (diphoton) production due to ADD gravity can happen only via $s$-channel graviton annihilation. It is typically studied in the central region (small \Dy)~\cite{ATLAS:2014gys, CMS:2014lcz, CMS:2018ipm, CMS:2018dqv, CMS:2018ucw, CMS:2018nlk, CMS:2021ctt}. ADD-based calculations often use linearized effective gravity, valid for $\mll(\mgg)\ll\md$. This regime is complementary to the trans-Planckian one, where the system mass greatly exceeds \md. The same argument applies to the production of high-\pt jet/lepton/photon + missing energy ($j$/$l$/$\gamma$ + ME) from real graviton emission.
 In the $\shat$-channel processes the energy is concentrated in a region of size $1/\sqshat$, leading to classical BH formation in the trans-Planckian regime $\sqshat \gg \md$. Subsequent BH evaporation produces leptons, photons, and jets with energies of the order of BH temperature  ($T\sim 1/ \shat^{\frac{1}{2(n_D+1)}}$). This yields high-multiplicity final states with energy distributed across many particles, rather than concentrated in dijet, dilepton, or diphoton systems \cite{Dimopoulos:2001hw}. So we expect the simultaneous signal of BH formation in the trans-Planckian ($\sqshat\gg\md$) regime but with $\that\sim\shat$. Current preliminary CMS results  exclude BH formation for masses below $\sim 11.4$ TeV at $\md=2$ TeV~\cite{CMS-PAS-EXO-24-028}.

To measure a cross section on the order of $\sim 10^{-6}$ pb, an integrated luminosity of at least $\sim1~\mathrm{ab}^{-1}$ is required, which is expected to be available at future facilities such as the HL-LHC, FCC-hh, and CEPC-SppC.  However, at high luminosities and large rapidities, it becomes necessary to disentangle overlapping \pp collisions occurring within the same or nearby bunch crossings. This imposes stringent requirements for high granularity in the next generation of detectors to effectively resolve and identify large rapidity jets from individual \pp collision events and correct for possible overlap effects.

\section{Summary}

The signal of gravity in the presence of large extra dimensions, as formulated by Arkani-Hamed, Dimopoulos, and Dvali (ADD)~\cite{Arkani-Hamed:1998jmv}, is studied for dijet production events with a large rapidity separation within the trans-Planckian eikonal regime defined by $\sqshat~\gg~\md~\gg~\sqrt{-\that}$ in proton-proton collisions. The calculations are performed for experimental conditions projected for future collider facilities such as the HL-LHC, FCCpp, and CEPC-SppC, covering \pp collision energies ranging from $\sqs = 13$~\TeV up to $100$~\TeV. Parameter values such as the number of extra dimensions $n_D = 2$ and $6$, and various Planck scales in presence of extra dimensions, \md, across collision energies are explored.

The QCD background is modeled using several approaches including LO+LL DGLAP, \PYTHIAeight with CP5 tune, and LL/NLL BFKL calculations, with the NLL BFKL calculation providing the most reliable background estimates. Sensitivity to ADD gravity is established for specific lower bounds of \md increasing with collision energy. The approach complements the existing LHC searches by probing a different kinematic regime and requires high integrated luminosities and high    detector granularity to resolve rare events at large rapidities.

The calculations also show that using DGLAP dynamics at large rapidities in the semi-hard regime may significantly overestimate the QCD which could result in misinterpreting experimental data and overlooking potential signals of new physics. This work demonstrates how the use of NLL BFKL background, rather than a DGLAP extrapolation, can alter the interpretation of dijet measurements with large rapidity separations.

Tabulated results are provided in the Zenodo record~\cite{egorov_2026_18540728}.

\begin{acknowledgments}

We thank the supercomputing system "Konstantinov" of the PIK Data Centre at NRC KI--PNPI for providing us with computing resources.

\end{acknowledgments}

\section*{Data availability}

The data that support the findings of this article are openly available~\cite{egorov_2026_18540728}.

\bibliography{transplanckian}

\end{document}